\documentclass{article}

\usepackage[utf8]{inputenc}
\usepackage{amsmath, amsfonts, dsfont}
\usepackage{graphicx}

\usepackage{etoolbox}
\newtoggle{oip}
\toggletrue{oip}
\togglefalse{oip}

\newtoggle{anon}
\toggletrue{anon}
\togglefalse{anon}

\iftoggle{oip}{
\usepackage[nolists]{endfloat}
}

\usepackage{longtable, booktabs}

\usepackage{natbib}

\usepackage{geometry}
\title{The Role of Emotions in Propagating Brands in Social Networks}

\iftoggle{anon}{

\author{}
\date{}

}{

\author{Ronald Hochreiter \and Christoph Waldhauser}
\date{September 2014}

}

\begin{document}

\maketitle

\iftoggle{oip}{

\subsection*{Abstract}

\noindent {\bf Purpose -- } A key aspect of word of mouth marketing are emotions. Emotions in texts
help propagating messages in conventional advertising. In word of mouth scenarios,
emotions help to engage consumers and incite to propagate the message further. While
the function of emotions in offline marketing in general and word of mouth marketing
in particular is rather well understood, online marketing can only offer a limited
view on the function of emotions. In this contribution we seek to close this gap.
\vskip 0.2cm
\noindent {\bf Design/methodology/approach -- } More than 30,000 brand marketing messages from the Google+ social networking site are collected. Using
state of the art computational linguistics classifiers, we compute the sentiment
of these messages. Starting out with Poisson regression-based baseline models,
we extend upon earlier research by computing multi-level mixed effects models that compare the
function of emotions across different industries.
\vskip 0.2cm
\noindent {\bf Findings -- } We find that while the well known notion of activating emotions propagating messages for marketing purposes holds in general for our data as well, there are significant differences between the observed industries.
\vskip 0.2cm
\noindent {\bf Practical implications -- } Despite its importance for marketing in general and brand management in particular, the precise mechanism of word of mouth message propagation in online social networks has not yet been adequately researched. We set out to fill this gap with additional insights, that will empower practitioners to better harness the power of word of mouth propagation across industries and help academia understand the different needs of consumers.
\vskip 0.2cm
\noindent {\bf Originality/value -- } To the best of our knowledge, this is the most comprehensive study of brand marketing in social networking sites in general and the first one using Google+ data. Our work's innovative value is rooted in shifting the focus from listening in on user streams to observing brand owners.
\vskip 0.2cm
\noindent {\bf Keywords} Marketing, Social media, Word of mouth, Propagation, Google+, Mixed effects, Emotion, Text mining, Computational linguistics
\vskip 0.2cm
\noindent {\bf Paper type} Research paper

\newpage

}{

\begin{abstract}
A key aspect of word of mouth marketing are emotions. Emotions in texts
help propagating messages in conventional advertising. In word of mouth scenarios,
emotions help to engage consumers and incite to propagate the message further. While
the function of emotions in offline marketing in general and word of mouth marketing
in particular is rather well understood, online marketing can only offer a limited
view on the function of emotions. In this contribution we seek to close this gap. We
therefore investigate how emotions function in social media. To do so, we collected
more than 30,000 brand marketing messages from the Google+ social networking site. Using
state of the art computational linguistics classifiers, we compute the sentiment
of these messages. Starting out with Poisson regression-based baseline models,
we seek to replicate earlier findings using this large data set. We extend upon
earlier research by computing multi-level mixed effects models that compare the
function of emotions across different industries. We find that while the well known
notion of activating emotions propagating messages holds in general for our data
as well. But there are significant differences between the observed industries.
\end{abstract}

\noindent {\bf Keywords} Marketing, Social media, Word of mouth, Propagation, Google+, Mixed effects, Emotion, Text mining, Computational linguistics

}
\section{Introduction}\label{introduction}

An important factor in marketing is the propagation of messages by word
of mouth. This lends the marketer's original message additional
credibility and extends its reach. This is an essential contribution in
the co-creation process of brands. The advent of social media and social
networking sites gave marketing a broader access to people's social
circle and their word of mouth message networks. Despite its importance
for marketing in general and brand management in particular, the precise
mechanism of word of mouth message propagation in online social networks
has not yet been adequately researched. In this paper we set out to fill
this gap with additional insights, that will empower practitioners to
better harness the power of word of mouth propagation across industries
and help academia understand the different needs of consumers.

In order to do so, we harvested a large number of social media marketing
messages. To the best of our knowledge, this is the most comprehensive
study of brand marketing in social networking sites in general and the
first one using Google+ data. We then computed their sentiment using
advanced state of the art computer linguistic classifiers. Starting from
well established Poisson-based regression models and extending to mixed
effects models, we seek to trace the precise function of sentiment in
marketing messages in the digital world. Due to the large number of
observations, the results are expected to be very reliable.

This paper is organized as follows. We first distill the
interdependencies between brand management, marketing and emotions in
Section \ref{marketing-brands-and-emotions}. Then we review other
contributions towards a better understanding of emotions in social
networks and word of mouth marketing in Section
\ref{word-of-mouth-in-social-networking}. In Section
\ref{research-design} we first describe our data and methods and then
the results of our analysis in the subsequent section. Finally, we
discuss our findings and offer some concluding remarks in Section
\ref{discussion-implications}.

\section{Marketing, Brands and
Emotions}\label{marketing-brands-and-emotions}

At the beginning of the 21st century, marketing theory took a sharp
turn. When conceptualizing marketing along the lines of the production
of goods, many phenomena remained unexplained: For instance, brands and
the involvement of consumers with brands could not be captured with a
production logic. In contrast, a service-dominant logic is much more
powerful in explaining especially the interactions between consumers and
brands \citep{merz2009evolving, vargo2004evolving, vargo2008service}.
There, consumers work with brand managers in concert to co-create a
brand. So instead of pure recipients of messages, consumers are
attributed with an active role in the marketing and communication
process \citep{hollenbeck2012consumers}. This active role revolves
around taking up messages that have been seeded by marketers and
propagate them further, with possible own additions to the original
message.

In essence, much of the co-creation process in marketing depends on word
of mouth \citep{moldovan2011different}. This is, the receiving and
taking up and altering of a message by consumers. \citet{holt2002brands}
describes that this also becomes one of the greatest challenges in the
process. Due to the distance between brand managers and consumers, a
lack of authenticity hinders much of the co-creation process. They find,
however, that with increased use of word of mouth, authenticity becomes
restored. This is also seconded by \citet{chu2011determinants} who find
that it is these connection properties that are intimately linked to
word of mouth effects.

There are two aspects to bolstering the building of brands with word of
mouth effects. One is the role of emotions, that in general are a very
powerful vehicle in marketing. The study by
\citet{stokburger2012drivers} identifies the memorable brand experiences
are one of the key drivers behind consumer--brand identification, and
emotions are triggering these experiences more easily
\citep{pham2013influence}. And the other one is the augmenting effect
that brand communities can have.

Strictly speaking, an emotion is a feeling or a sensation that is a
process within a human being. This narrow definition of emotion allows
emotions to be triggered by external influences \citep{bagozzi1999role}.
While this distinction is perhaps worthwhile for psychological analyses,
widening this definition, would apply the term emotion also to the
triggering influence. This allows a more natural interpretation of
emotions in texts. For instance a text that \emph{is} sad will trigger
the emotion of sadness in its readers. We use the term emotion in this
latter, wider sense.

Emotions have various effects on the way marketing messages are
received. In a word of mouth marketing context, for instance
\citet{berger2012makes} found that using the right emotions increases
the likelihood of virality of a message. In their work they conclude
that high-arousal emotions like awe and anger will increase the virality
of a message. Less activating emotions, like sadness, work against the
propagation of messages. \citet{romani2012emotions} arrive at different
conclusions when diagnosing that anger and sadness as emotions in
marketing messages will trigger negative word of mouth events.

Therefore, emotions in brand messages affect the degree of consumer
involvement in the co-creation process.

An area that is or should be of elevated interest to marketers are brand
communities \citep{schau2009brand}. Brand communities are groups of
people that form around brands. They are not necessarily geographically
close and might share nothing more than their common interest in a brand
\citep{muniz2001brand}. Obviously, due to the centrality of a brand in
these communities, they play an important role in the co-creation of the
brand \citep{kozinets2010networked}.

\section{Word of Mouth in Social
Networking}\label{word-of-mouth-in-social-networking}

The advent of social media plays along well with the renewed
understanding of marketing as omnidirectional co-creating process
\citep{onishi2012marketing, kunz2011toward}. Instead of simply selling
news to the masses \citep{iyengar2010news}, social media allows for
users selecting messages to propagate themselves. This makes social
media ideal for word of mouth based marketing. Social networking sites
(like Facebook) are a special case of social media that emphasizes
shortness and multimedia over more elaborate texts.

When translating the mechanics of word of mouth marketing to the social
networking arena, some equivalents become apparent
\citep{matteo2013branding}. The propagation of a message can be achieved
by using one of two forms of endorsement: a lesser and a more potent
one. The details of these forms vary by implementation, but there is a
general pattern. Each form of endorsement leads to the original message
being injected into the news stream of users connected to the endorsing
user. The lesser form of endorsement (Facebook's likes, Twitter's stars
and Google's +1s) leads to a less prominent injection and does not allow
to enrich the original message with own content. The second form of
endorsement (Reshares or Retweets) prominently adds the original story
(possibly with additional information) to the users news streams.

The brand community has multiple equivalents. One is the official
company sponsored social networking presence. There users can meet to
retrieve seeded brand messages directly from the source. They can use
these messages to propagate them further down their own stream of
followers, possibly enriching it with their own respective contents on
the way. Other brand community analogues are less formal communities or
unofficial fan pages.

While word of mouth based marketing can be readily implemented using
social media, in the past there have also been some challenges to
companies discovered. \citet{onishi2012marketing} find that social media
and traditional mass media marketing can be hindering each other if not
planned well. Only by taking into consideration the particularities of
each channel can an optimal outcome be secured.

However, the greatest challenge perhaps is, that in social media in
general but social networking in particular, brand managers are much
harder pressed to anticipate what consumers want
\citep{arango2013social}. This is mainly because consumers can (and
will) react instantly to seeds planted by the brand manager. Because of
the nature of social media, these reactions can be far reaching, indeed.
Take for instance the by now canonical example of the brand of a media
manager that she managed to destroy during a flight from the US to
Africa by posting a single negative message.

It is therefore crucial for brand managers to understand which messages
lead to increased resharing of her brand messages. As stated above, the
equivalent of word of mouth in social media is the two forms of
endorsement. We have also established the notion, that emotions are
viable predictors for the virality of social media messages. However,
most current research is focused on single industries. For instance,
both \citet{berger2012makes} and \citet{chevalier2006effect} focus on
the book publishing industry. While informative, their research is hard
to generalize, given their limited sample. The results described above
draw an inconsistent picture with the role of emotions still being
fuzzy. This is potentially due to the same emotions functioning
differently across industries.

In this paper we want to focus on differences between industries. In
this paper we focus on two key aspects of message sentiment and their
effects on message endorsements: 1. How does message polarity affect
message endorsements? 2. How does the use of emotions affect message
endorsements?

In sight of the current state of the art, we offer the following
hypotheses:

\begin{enumerate}
\def\labelenumi{\arabic{enumi}.}
\itemsep1pt\parskip0pt\parsep0pt
\item
  Messages with a positive polarity are more likely to be endorsed.
\item
  The use of (right) emotions in social media marketing messages
  increases the count of endorsements that message receives.
\item
  Not all emotions have the same effect across industries.
\end{enumerate}

In the following section we are going to discuss both method and data we
used to answer this question.

\section{Research Design}\label{research-design}

In order to generate a complete picture of the determinants of emotional
brand marketing in social media, a large host of brands needs to be
analyzed. We used the Open Knowledge Foundation's Brand Repository data
base \citep{okfn}. At the time of analysis, that data base contained the
names, identifiers and websites of 4151 different brands. To ensure
comparability, we reduced this list further to contain only brands that
used a .com top-level domain. This effectively excludes local and
localized brands, focusing on internationally available brands. This
step also excluded brands that did not use digital marketing at all, as
demonstrated by their lack of a brand website. We further excluded any
brands that either did not have a Google+ profile at all or one with a
very limited +1 count ($p1 < 250$). The resulting list of 199 brands was
then manually checked and each brand assigned to an industry. During the
manual check and classification, we removed another 50 brands from the
list that either had not posted anything on Google+ or that had slipped
through our selection heuristic. Because brands from the automotive
industry were somewhat underrepresented, we added nine randomly chosen
brands to the list. In the end, 156 brands remained.

For each brand, all the posts to their Google+ page were retrieved using
R \citep{Rcore} and the \emph{plusser} extension package
\citep{plusser}. This yielded 32409 posts in total. Table
\ref{tab:indBrands} gives the distribution of brands and posts over
industries.

\begin{table}[ht]
\centering
\begin{tabular}{rlrr}
  \hline
 & Industry & Pages & Posts \\ 
  \hline
1 & Apparel &  31 & 6368 \\ 
  2 & Automotive &  11 & 2200 \\ 
  3 & Cosmetics &  37 & 7742 \\ 
  4 & Electronics &  14 & 2985 \\ 
  5 & Food/Beverages &  43 & 8927 \\ 
  6 & Sports &   8 & 1600 \\ 
  7 & Other &  12 & 2587 \\ 
   \hline
\end{tabular}
\caption{Number of analyzed brands and posts per industry.} 
\label{tab:indBrands}
\end{table}

Using the naive Bayes classifier implemented in the sentiment package
\citep{sentiment}, the messages' polarity and emotionality were
computed. The package delivers a negative-positive ratio to measure
polarity based on keywords that occur in the message. To arrive at an
estimation of sentiment, the package uses a similar approach for the
emotions \emph{Anger}, \emph{Disgust}, \emph{Fear}, \emph{Joy},
\emph{Sadness}, and \emph{Surprise}.

As stated above, the key concept we seek to analyze is the propagation
of a message in social networks. The analogue of word of mouth
propagation in the online scenario is the higher form of endorsement:
reshares. In order to be able to analyze the and compare the
distribution of reshares, we built two types of linear models. One type
to establish a Google+ baseline that is interesting in itself, as
Google+ has not yet received widespread academic attention, is a simple
generalized linear model (GLM). We chose a GLM because the reshare
counts of messages cannot be considered to be normally distributed.
Rather, they represent rare (given the number of followers) events that
are best modeled using a Poisson process. GLMs permit just that. The
following equation summarizes our model.

\[ r \sim p + \mathbf{e} + \mathbf{cv} + \mathrm{offset(t)} + \mathrm{offset(f)}\]

On the left hand side, $r$ represents the number of reshares a message
has received. The variables $p$ and $\mathbf{e}$ contain the polarity
ratio and the emotions encoded within a message in six dimensions,
respectively. There are other determinants that might influence reshare
frequency independent to our analytical problem. Among those are message
length and time of day the post was put online
\citep{stieglitz2012political, cha2010measuring, suh2010want, yang2010predicting, hochreiter2013stochastic}.
Among other covariates that have been identified to be relevant for
Twitter, but are not implemented in the Google+ search API are
e.g.~hashtags and media attachments. Message length and time of day are
contained in $\mathbf{cv}$.

Poisson based regression assumes a constant window of measurement. In
the case at hand, the window's dimensions are given by the age of a post
($t$) and the number of followers it has been exposed to ($f$). While
the age of the post can be reliably determined, we assume the number of
page followers to have remained constant over the period of
investigation and therefore take this figure as the number of followers
the post has been exposed to. Both terms need to be included as offsets
in the model to adequately model the relationships between the number of
reshares and the message properties as described above.

This simple model is then extended to allow for brand and industry
comparisons by using mixed effects models. Conceptually, we consider our
selection of brands and industries to be random. Therefore, we model any
effects due to industry and brand as random effects. We test both random
intercept and random slope models against each other.\footnote{All
  modelling was done using the lme4 package \citep{lme}; graphics were
  produced using ggplot2 \citep{ggplot}.}

In this section we described our method of retrieving posts from Google+
and computing the posts' sentiments. We proposed Poisson regression
models to use sentiment to explain the reshare count of a message. In
order to incorporate company and industry differences, we use mixed
effect models. The results of our analyses will be given in the next
section.

\section{Results}\label{results}

In order to establish a baseline of message propagation mechanisms in
Google+, we initially computed simple Poisson regression models. We
started out with an empty model and included it as a reference ($m_0$).
Next, we added variables to control for the message length (number of
characters) and time of day (deviance from noon in hours). This formed
model $m_1$. In a next step, we controlled for the number of comments
and the number ob +1s a message has received in $m_2$. Finally, in $m_3$
and $m_4$, the polarity and emotionality predictors were added. All
models were tested successively against each other using standard
Likelihood Ratio tests. All test yielded highly significant results with
$p<2.2 \times 10^{-16}$. This is to be expected given the large sample
size. More appropriately, we compared the models using the AIC
criterion. The results from this baseline estimation are given in Table
\ref{tab:glm}.

\begin{table}[ht]
\centering
\begin{tabular}{rrrrrr}
  \hline
 & $m_0$ & $m_1$ & $m_2$ & $m_3$ & $m_4$ \\ 
  \hline
Intercept & 0.00 & 0.00 & 0.00 & 0.00 & 0.00 \\ 
  Message length &  & 1.00 & 1.00 & 1.00 & 1.00 \\ 
  Time of Day &  & 1.04 & 1.01 & 1.01 & 1.01 \\ 
  Comments &  &  & 1.01 & 1.01 & 1.01 \\ 
  +1s &  &  & 1.00 & 1.00 & 1.00 \\ 
  Polarity &  &  &  & 1.00 & 1.00 \\ 
  Anger &  &  &  &  & 1.05 \\ 
  Disgust &  &  &  &  & 0.89 \\ 
  Fear &  &  &  &  & 0.76 \\ 
  Joy &  &  &  &  & 0.95 \\ 
  Sadness &  &  &  &  & 0.95 \\ 
  Surprise &  &  &  &  & 1.02 \\ 
  AIC & 926202.56 & 921079.66 & 794331.11 & 793457.15 & 786504.72 \\ 
   \hline
\end{tabular}
\caption{Comparison of simple Poisson-based GLMs. Incidence Rate Ratios and model AIC.} 
\label{tab:glm}
\end{table}

In order to analyze and compare the effects of emotions on message
propagation across industries, we computed multi-level models. There,
messages are nested within industries. We compared two kinds of model
types, all containing the same baseline fixed effects as $m_4$ above and
one sentiment. We tested both model types for all seven sentiments
(polarity and six emotions) using likelihood ratio tests. All tests
produced overwhelming evidence that random slope models were required.
Table \ref{tab:eff} contains the incidence rate ratios that were
computed using the models' random effects into account for the different
industries in the data set. The table's $\sigma$ line describes the
standard deviation of the random effects. The larger the value becomes,
the more heterogeneous the function of this sentiment is across the
surveyed industries.

\begin{table}[ht]
\centering
\begin{tabular}{rrrrrrrr}
  \hline
 & Polarity & Anger & Disgust & Fear & Joy & Sadness & Surprise \\ 
  \hline
Apparel & 1.01 & 0.86 & 0.73 & 0.63 & 0.94 & 0.96 & 0.98 \\ 
  Automotive & 1.00 & 1.14 & 1.75 & 1.12 & 1.01 & 1.05 & 1.02 \\ 
  Cosmetics & 1.00 & 0.96 & 0.88 & 0.83 & 1.02 & 1.05 & 1.08 \\ 
  Electronics & 1.00 & 0.55 & 0.74 & 0.77 & 0.83 & 1.05 & 0.98 \\ 
  Food/Beverages & 1.01 & 0.73 & 1.20 & 0.72 & 1.00 & 0.96 & 1.24 \\ 
  Sports & 1.00 & 0.94 & 1.94 & 0.96 & 1.03 & 1.05 & 0.99 \\ 
  Other & 1.02 & 1.04 & 0.52 & 0.85 & 1.07 & 0.82 & 0.90 \\ 
  $\sigma$ & 0.01 & 0.28 & 0.30 & 0.26 & 0.07 & 0.09 & 0.09 \\ 
   \hline
\end{tabular}
\caption{Varying functions of emotions as indicated by incidence rate ratios for message sentiment.} 
\label{tab:eff}
\end{table}

In this section we first established a base line for the function of
sentiments in Google+ posts in the context of word of mouth marketing.
We could validate previous findings for other forms of social media in
the importance of covariates. We then proceeded with including random
effects to model different industries. Using hypothesis testing, we
conclude that more complex random slope models are required. In the next
section we are going to discuss the implications of these findings.

\section{Discussion \& Implications}\label{discussion-implications}

The results exhibited above allow for two important insights. One is
that Google+ does not behave differently from other social networks, in
terms of covariates explaining the likelihood of message propagation.
The other key insight are the differences between industries in the
function of sentiments.

\subsection{Baseline}\label{baseline}

Looking at the results from $m_4$, it becomes evident that the findings
by \citet{berger2012makes} hold also for Google+. When controlling for
message length and time of day, activating emotions like \emph{Anger} or
\emph{Surprise} exhibit incidence rate ratios larger than 1. This
indicates that every increase in \emph{Anger} also leads to an increase
in the likelihood of the message being reshared. As expected, the
\emph{Sadness} emotion is different. There, an increase in
\emph{Sadness} leads to a decrease in expected reshares. Due to the
large number of brands and messages sampled and all results being
significant, these results allow for a high degree of confidence.

\subsection{Industry comparison}\label{industry-comparison}

Turning to the comparison of the function of sentiments across
industries reveals a different picture. When discussing random effects
models, there are two key figures. One is the variability observed and
the other one the random effects computed.

\begin{figure}
\begin{center}
\includegraphics[scale=0.6]{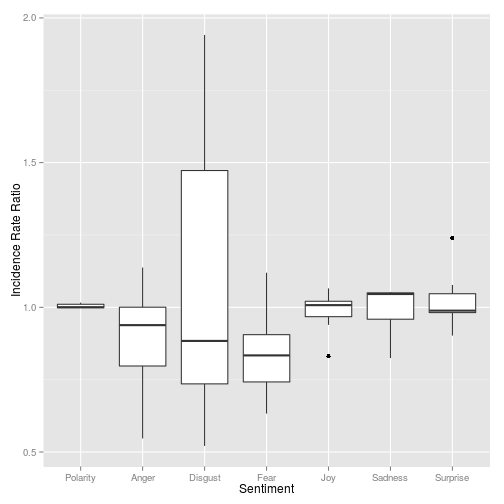}
\caption{Boxplot comparing the variablity in sentiment effect accross industries.}
\label{fig:boxp}
\end{center}
\end{figure}

Figure \ref{fig:boxp} compares the variability in all seven sentiments
of the industries observed. \emph{Disgust} exhibits a very large
variability, closely followed by \emph{Fear} and \emph{Anger}. The
remaining emotions and \emph{Polarity} appear to have almost constant
effects across industries. While there is very little variation in
\emph{Joy} but also in \emph{Sadness} and \emph{Surprise}, negative but
activating emotions like \emph{Anger}, \emph{Disgust} and \emph{Fear}
differ radically in their function across industries.

\begin{figure}
\begin{center}
\includegraphics[scale=0.8]{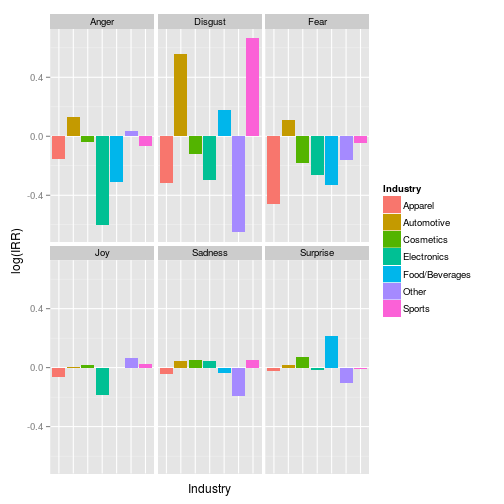}
\caption{Effect of emotion on message propagation likelihood accross industries.}
\label{fig:barch}
\end{center}
\end{figure}

Figure \ref{fig:barch} displays the different effects the measured
emotions have on message propagation for brands of different industries.
The high variability of \emph{Disgust} is also clearly visible in its
panel in this figure, while \emph{Joy} and \emph{Sadness} remain fairly
constant. Looking at \emph{Anger}, we find that most random effects
indicate a negative effect on message propagation. Only the automotive
industry can harness \emph{Anger} successfully. The emotion of
\emph{Disgust} has some strong propagating effects also in the
automotive and the sports industries. \emph{Fear} is an emotion that is
unpopular with consumers of goods from most industries but particularly
apparel.

The findings of \citet{berger2012makes} would suggest that \emph{Anger}
contributes to and \emph{Sadness} hinders message propagation. We can
confirm the function of \emph{Sadness} in our data. For almost all
industries, \emph{Sadness} in the message leads to a decrease in the
likelihood of a message being reshared. While brands from the
automotive, cosmetics and electronics industries enjoy a slightly
positive effect of \emph{Sadness}, it is so small, that it can safely be
ignored. This, however, is at odds with the results provided by
\citet{romani2012emotions}, that would expect sadness to lead to
increased word of mouth effects. Our data does not support this.

\emph{Anger} on the other hand does not behave as the predicted word of
mouth driver. It is only the automotive industry that can benefit
clearly from an angry emotion in its messages. There are barely
noticeable effects for cosmetics and sports. Anger, however, clearly
hinders message promotion in the industries of Electronics and
Food/Beverages.

Recurring to the research questions we posited in the beginning, we have
to conclude that message polarity does not only not vary across
industries. Message polarity does not influence the likelihood of
message propagation at all. There are, therefore, no connections between
the polarity of a message and the strong form of endorsement.

Turning to the second question, it is apparent that there are stark
differences in the functioning of emotions between the brands of
different industries. The perhaps most important finding lies here:
activating emotions do not function in the expected way for a majority
of industries.

This has serious managerial implications. While traditional word of
mouth based marketing literature asserts that activating emotions will
\emph{always} support the propagation of word of mouth, we find that
this is only the case for select industries. Brand managers seeking to
extend their involvement in the co-creation process are required to
tread carefully in the light of these findings. While \emph{Anger} will
help in getting messages adopted by consumers for the automotive
industry, managers responsible for electronics brands will need to avoid
this emotion in their messages.

In the last instance, it appears that the brand co-creation process is
more complicated than previously thought. With clear and significant
differences in the function of sentiments between industries, brands and
their advertisements remains a game for highly skilled players and
careful considerations.
\bibliographystyle{plainnat}
\bibliography{lit}

\iftoggle{oip}{

\iftoggle{anon}{}{

\subsection*{About the authors}

Ronald Hochreiter is Docent and Assistant Professor at the Department of Finance, Accounting and Statistics at the WU Vienna University of Economics and Business. His main interests include Data Science in general and Optimization under Uncertainty in particular. He holds a PhD degree in Computational Management Science from the University of Vienna and has published more than 40 articles in international journals, conference proceedings and books.

Christoph Waldhauser is PhD candidate at the Department of Methods in the Social Sciences at Vienna University. As a data scientist, he has worked intensively on social media marketing and evolutionary optimization. Christoph is the corresponding author and can be contacted at: chw@kdss.at

}

}

\end{document}